\documentstyle[11pt,aasms4]{article}
\input epsf

\begin{document}

\title{PLASMA MODES ALONG THE OPEN FIELD LINES OF A NEUTRON STAR}
\author{U. A. Mofiz}
\affil{BRAC University, 66 Mohakhali, Dhaka - 1212, Bangladesh;\\
mofiz@bracuniversity.ac.bd} \and
\author{B. J. Ahmedov}
\affil{Institute of Nuclear Physics and Ulugh Beg Astronomical Institute,
Astronomicheskaya 33, Tashkent-100052, Uzbekistan; \\
ahmedov@astrin.uzsci.net}

\begin{abstract}

We consider electrostatic plasma modes along the open field lines
of a rotating neutron star. Goldreich-Julian charge density in
general relativity is analyzed for the neutron star with zero
inclination.  It is found that the charge density is maximum at
the polar cap and it remains almost same in certain extended
region of the pole. For a steady state Goldreich-Julian charge
density we found the usual plasma oscillation along the field
lines; plasma frequency resembles to the gravitational redshift
close to the Schwarzschild radius. We study the nonlinear plasma
mode along the field lines. From the system of equations under
general relativity, a second order differential equation is
derived. The equation contains a term which describes the growing
plasma modes near Schwarzschild radius in a black hole
environment. The term vanishes with the distance far away from the
gravitating object. For initially zero potential and field on the
surface of a neutron star, Goldreich-Julian charge density is
found to create the plasma mode, which is enhanced and propagates
almost without damping along the open field lines. We briefly
outline our plan to extend the work for studying soliton
propagation along the open field lines of strongly gravitating
objects.

\end{abstract}


\section{INTRODUCTION}

Study of plasma modes in the neutron star or black hole
environments is related with the investigation of radio emissions
coming from these sources (see, e.g.  Buzzi et al. 1995, Mofiz
1997 and the references therein). Radio pulsars which are rotating
neutron stars with spin periods ranging from  ms to 5 s, are
characterized by surface magnetic fields of the order of $10^{12}$
G, radii of about 10 km and central densities in excess of
$10^{14}$ $g\cdot cm^3$, and so are purely gravitating objects. A
spining magnetized neutron star generates huge potential
differences between different parts of its surface (Goldreich $\&$
Julian 1969 ). The cascade generation of electron-positron plasmas
in the polar cap region (Sturrock 1971, Ruderman $\&$ Sutherland
1975) means that the magnetosphere of a neutron star is filled
with plasma - screening the longitudinal electric field. This
screening results in the corotation of plasma with a star. Such a
rotation is not possible outside the light cylinder, thus it forms
essentially different groups of field lines: closed i. e. those
returning the stellar surface, and open, i.e. those crossing the
light cylinder and going to infinity.  As a result, plasma may
leave the neutron star along the open field lines. The charges
along the field lines create plasma modes which may be related
with the pulsar radiation and with its microstructures.

Our study of plasma modes along the field lines is boosted by the
pioneering works of Goldreich $\&$ Julian (1969), Sturrock (1971), Mestel
(1971), Ruderman $\&$ Sutherland (1975) and Arons $\&$ Scharleman (1979).
The subsequent achievements and some new ideas are reviewed  by Arons
(1991), Michel (1991), Mestel (1992) and Muslimov $\&$ Harding (1997).
Although a self consistent pulsar magnetosphere theory is yet to
developed, the analysis of plasma modes in the pulsar magnetosphere based
on the above mentioned works provides firm grounds for the construction of
such a model.

In this paper, we attempt to extend Muslimov $\&$ Harding work
(1997) to study plasma modes along the open field lines of a
rotating neutron star. In $\S2$ general relativistic equations
describing the electrodynamics of a rotating neutron star are
formulated. The equations are rewritten in the frame of reference
corotating with the neutron star.  We deduce the general system of
equations governing the electrostatic modes in the pulsar
magnetosphere.  A detail analysis of Goldreich-Julian charge
density in general relativity in done in $\S3$.  It is shown that
the charge density exponentially decays with the distance away
from the surface of the star while it has a periodical dependence
on the polar angle along the surface. The field is maximum at the
polar cap region and it remains almost same in certain extended
region in the pole. In $\S4$ we study the linear plasma modes
along the open field lines. A general equation governing
electrostatic potential is derived. For a steady state
Goldreich-Julian charge density, the usual plasma oscillation
along the field lines is found. Plasma frequency resembles to the
gravitational redshift close to Schwarzschild radius while at a
large distance from the gravitational radius, it is the usual
plasma oscillation along the field lines. In $\S5$ we study the
nonlinear plasma modes along the field lines. From the system
equations under general relativity, a second order differential
equation is derived. The equation contains a term which describes
the growing plasma mode near the Schwarzschild radius of a neutron
star or a black hole. The term vanishes with the distance far away
of the gravitating object. The equation is solved numerically
subjected to appropriate boundary conditions. It is found that
Goldreich-Julian charge density creates the initial field on the
surface of the star which is enhanced near the gravitational
radius and almost without damping propagates along the open field
lines. In $\S6$ we conclude our findings and discuss them for
further investigations.

\section
{GENERAL RELATIVISTIC ELECTRODYNAMIC EQUATIONS IN THE COROTATING FRAME
OF REFERENCE}

Recently Muslimov and Harding (1997)
derived the general relativistic electrodynamic equations for a neutron
star in the corotating frame of reference. It is noted that the effects of
general relativity are very important : the dragging of inertial frames of
reference significantly affects the electric field generated in the
vicinity of a rotating magnetized neutron star, while the static part of
the gravitational field results in additional enhancement of electric and
magnetic fields near a star.

The metric of an asymptotically flat, stationary, axially
symmetric spacetime around a rotating gravitating body (see, e.g.
Landau $\&$ Lifshitz 1975) is considered. In spherical polar
coordinates $ x^0 = ct, x^1 = r, x^2 = \theta $ and $x^3=\phi,$ we
have
$$ds^2 = A^2 (cdt) ^2 -B^2(dr)^2-C^2(d\theta)^2-D^2(d\phi-\omega dt)^2,
\eqno (1)$$
where $A = B^{-1} = ( 1-{r_g/ r})^{1/2}$ is the gravitational
redshift function, $C=r, D= r\sin\theta$,  $r_g=2GM/c^2$ is the
gravitational radius of body (neutron star) of mass M, J is the
angular momentum of a neutron star, c is the speed of light and G
is the gravitational constant. The metric in equation (1) is the
approximation of Kerr metric when the ratio $J/Mcr_g$ is small.
The presence of the nondiagonal component in metric in equation
(1)  results in the well known effect of dragging of inertial
frames of reference (the Lense-Thirring effect)  with the angular
velocity

$$\omega={2GJ\over c^2r^3}.  \eqno(2)$$

The metric in equation (1) can be transformed to the frame of
reference corotating with a neutron star:
$$ds^2 = A^2 (cdt) ^2 -B^2(dr)^2-c^2(d\theta)^2-D^2(d\phi-\Omega dt)^2,
\eqno (3)$$
by transformations $t^{{\prime }}=t, r^{^{\prime }}=r, \theta
^{^{\prime }}=\theta , \varphi^{^{\prime }}=\varphi -\Omega_0 t$.
Here $\Omega=\omega -\Omega_0$, where $\Omega_0$ is the angular
velocity of rotation of star relative to the distant observer.

The zero angular momentum observer (ZAMO; see, e.g. Thorne, Price
\& Macdonald 1986)
 has the four-velocity
$$e^\nu\{\frac{1}{\sqrt{1-r_g/r}},0,0,-\frac{\Omega}{c\sqrt{1-r_g
/r}}\};\qquad e_\nu\{-\sqrt{1-r_g/r},0,0,0\}\ .\eqno(4)$$

Then the general-relativistic Maxwell equations for observer
(eq.[4]) in the metric (eq.[3]) take the form
$$ \mathbf{\nabla}\cdot {\bf B} =0\ , \eqno(5a) $$

 $$\mathbf{\nabla}\times\left(\alpha {\bf E} - \left({\mathbf{ \omega -
\Omega_0}}\right)\times \mathbf{B}\right)= -{1\over
c}\frac{\partial{\bf B}}{\partial t}, \eqno(5b)$$

$$\mathbf{\nabla}\cdot{\bf E} = 4\pi\rho, \eqno(5c)$$

$$ \mathbf{\nabla}\times \left(\alpha{\bf B}\right)
+\mathbf{\nabla}\times\left(\left(\mathbf{\omega-\Omega_0}\right)
\times {\bf E}\right) - \left(\mathbf{\omega-\Omega_0}\right)
\left(\mathbf{\nabla}\cdot{\bf E}\right)= {1\over c}{\partial {\bf
E}\over
\partial t} + {4\pi\over c}\alpha{\bf j}\ , \eqno (5d)$$
where $({\mathbf\Omega\times
B})^\alpha=e^{\alpha\beta\mu\nu}e_\beta\Omega _\mu B_\nu$,
${(\mathbf\omega-\Omega_0)}_\alpha=\left\{0,0,0,\left[\left({\omega
- \Omega_0}\right)r\sin\theta\right]/{c\sqrt{1-r_g/r}}\right\}$.

 Similarly, we may write the charge continuity equation in the above
mentioned frame as

 $$ {\partial\rho\over{\partial t}} + \left({\kappa\over{r^3}} -1\right)
\Omega_0\mathbf{m}\cdot \mathbf{\nabla}\rho +
 \mathbf{\nabla}\cdot{(\alpha\bf j)}=0. \eqno(6)$$

 Finally, the equation of motion of a charged particle is
$${1\over\alpha}{ d{\bf p}\over dt} = m\gamma {\bf g} + q \left({\bf E} +{{ \bf
v}\over c} \times {\bf B}\right)+ {\bf f}\ . \eqno(7)$$
Here $\alpha = (1-{r_g/ r})^{1/2}$ ($\equiv A $, as denoted in the
metric in equation (1) ), the
 parameter $\kappa\equiv {{2r_{g}R^2}/ 5}$, $R$ is the radius a
neutron star, $\rho = \sum_{s} n_{s}q_{s}, {\bf j}= \sum_{s} n_{s}q_{s}\bf
v_{s}$, $\bf v_s$ is the velocity, $q_{s}$ is the charge of particle,
$n_{s}$ is the particle number density, and summation is over all the
species; ${\bf p}= m\gamma{\bf v} $ is the momentum of the particle,
 $\gamma=(1-{v^2/ c^2})^{-1/2}$ is the Lorentz factor, $m$ is the rest
mass of the particle, $\bf f$ is an external force other than
electromagnetic, and $\bf g$ is the
 gravitational acceleration.  All electrodynamic quantities as magnetic
field $\mathbf B$, electric field $\mathbf E$, conduction current
$\mathbf j$, and charge density $\rho$ in these equations are such
as measured by ZAMO (eq.[4]). Gradient, curl and divergence are
taken along the curvilinear coordinate
\begin{equation}
e_{\hat{r}}=Ae_{r}=A\frac{\partial}{\partial r}\ , \quad
e_{\hat{\theta}}=\frac{1}{r}e_{\theta}=\frac{1}{r}\frac{\partial}{\partial
\theta}\ , \quad
e_{\hat{\phi}}=\frac{1}{r\sin\theta}e_{\phi}=\frac{1}{r\sin\theta}\frac{\partial}{\partial
\phi}\ , \nonumber
\end{equation}
$\mathbf{m}=r\sin\theta e_{\phi}$ is the Killing vector,
responsible for axial symmetry.

 Assuming  ${{\partial {\bf B}}\over{\partial
t}}=0$ in equation (5b) (i.e. considering magnetic field
 of a neutron star is stationary in the corotating frame),
from equation (5b) we get
 $$\alpha{\bf E}-({\bf\omega-\Omega_0)\times B}=-\mathbf{\nabla}\Phi,
\eqno(8)$$
where $\Phi$ is a scalar electrostatic potential.

 Taking the divergence of equation (8) and making use of equation (5c), we
get
 $$\mathbf{\nabla}\cdot\left({1\over\alpha}\mathbf{\nabla}\Phi + {1\over{\alpha }}({\mathbf
\Omega_0-\omega)\times B}\right)
 =-4\pi\rho\ .  \eqno(9)$$

 Equation (9) can be written as (see Muslimov $\&$ Tsygan 1986; Beskin 1990)

 $$\mathbf{\nabla}\cdot ({1\over \alpha}\mathbf{\nabla}\Phi)=-4\pi(\rho-\rho_{GJ})
\eqno(10)$$
 where
 $$\rho_{GJ}=-{ 1\over{4\pi }}\mathbf{\nabla}\cdot\left({1\over\alpha}({\bf
\Omega_0- \omega})\times
 {\bf B}\right)= -{1\over{4\pi }}\mathbf{\nabla}\cdot\left({1\over{\alpha}}(1-{
 \kappa\over{r^3}}){\mathbf\Omega_0\times B}\right) \eqno(11)$$

 is the relativistic analog of the Goldreich-Julian (1969) charge density.

 Finally, the equation of motion of the charged particle is
 $$\left[{\partial\over{\partial t}} + \left({\kappa\over{r^3}} -1\right){\mathbf
\Omega_0}\cdot\mathbf{\nabla} +\alpha{\bf
v}\cdot{\mathbf{\nabla}}\right]{\bf p} =-q {\mathbf{\nabla}}\Phi,
\eqno(12)$$
where the gravitational acceleration {\bf g} and the nonelectromagnetic force
$\bf f$ are justifiably ignored.

\section{GOLDREICH-JULIAN CHARGE DENSITY IN GENERAL RELATIVITY}

 In a pioneering work, Goldreich $\&$ Julian (1969) have shown that a
strongly magnetized, highly conducting neutron star, rotating
about the magnetic axis, would spontaneously build up a charged
magnetosphere. The essence of the argument is that it imposes a
charge magnetosphere which are subject to enormous unbalanced
electric forces parallel to the magnetic field. Goldreich and
Julian hypothesized that a far better approximation for the
magnetosphere would be shorting-out of the component of $\bf E$
along $\bf B$ by charges originating in the star. The
magnetospheric charges that maintain ${\bf E\cdot B} =0 $ are
themselves subject to the $\bf E\times B$ drift which sets them
into corotation with the star. Here, we analyze Goldreich-Julian
charge density in general relativity.

 Assuming zero inclination of the rotating star with the magnetic axis we
consider ${\bf B} =\{B_{r},B_{\theta},0\}$. The components $B_{r},
B_{\theta}$ in this case were first derived by Ginzburg \& Ozernoy
(1964). Later on, similar expressions were derived in a number of
papers (see, e.g.  Wasserman $\&$ Shapiro 1983; Muslimov$\&$
Tsygan 1986):

 $$ B_{r}={{2\cos\theta}\over{r^3}}f(r)\mu, \eqno(13a)$$

 $$B_{\theta}={{\sin\theta}\over{r^3}}\psi(r)\mu, \eqno(13b)$$
 where

 $$f(r)=-{{3r^3}\over{r_{g}^3}}\left[\ln\left(1-{r_{g}\over r}\right)+{{r_{g}}\over r}
+{1\over 2} \left({{r_{g}}\over r}\right)^2 \right],  \eqno (14)$$

 $$\psi (r)={{ 3r^2}\over{r_{g}^2}}\left[ {1\over{1-{{r_{g}}\over r}}} +
 2 {r\over{r_{g}}}\ln\left(1-{r_{g}\over r}\right) +1 \right]\sqrt{1-{{r_{g}}\over
r}}, \eqno(15)$$
 and $\mu$ is the magnetic dipole moment of a neutron star.

 We perform a detail calculation of $\rho_{GJ}$ with the magnetic field of
 a rotating neutron star given by equations (13a) and (13b), respectively.
 From equation (11), we find

 $$\rho_{GJ}=-{{\Omega_0}\over{4\pi c r^2\sin\theta}}\left\{
\left[\left(1-{\kappa\over{r^3}}\right)\frac{
 r^3\sin^2\theta B_{\theta}}{\sqrt{1-r_g/r}}\right],_{r} -
\left[\left(1-{\kappa\over{r^3}}\right)\frac{r^2\sin^2\theta
B_{r}}{1-r_g/r}\right],_{\theta}\right\}. \eqno(16)$$

  The calculation shows that

  $$\rho_{GJ}(r,\theta)=\frac{3\Omega_0 \mu}{4\pi c r^3}
  \left[ F_{1}(\bar r)\sin^2\theta - F_{2} (\bar r) (\sin^2\theta-2
\cos^2\theta)\right], \eqno(17)$$
  with
   $$F_{1}(\bar r)={\bar r}^3\Bigg\{( (1-{\beta\over{
\bar r^3}})\left\{\frac{2}{\bar r-1}-\frac{1}{(\bar r-1)^2}
+2\ln(1-\frac{1}{\bar r})\right\}$$
 $$+(2+\frac{\beta}{\bar r^3})\left\{{1\over \bar r}+\frac{1}{\bar r-1}
+2\ln (1-\frac{1}{\bar r})\right\}\Bigg\}, \eqno(18)$$

$$F_{2}(r)={\bar r}^3\frac{2(1-\frac{\beta}{\bar r^3})}{1-\frac{1}{\bar r}}
\left\{\frac{1}{2\bar r^2} +\frac{1}{\bar r}+ \ln(1-\frac{1}{\bar
r})\right\}. \eqno(19) \setcounter{equation}{19}
$$
  Here, $\bar r= {r/{r_{g}}}$ and $\beta= {\kappa/{r_{g}^3}}.$
  Asymptotically as $r/r_{g}\rightarrow \infty$ functions $F_1(\bar r)$ and
   $F_2(\bar r)
  \rightarrow 1$.

Thus Goldreich-Julian space charge has two purely general
relativistic contributions, one is due to the Schwarzschild
gravitoelectric parameter $r_g$ and the second one is due to the
gravitomagnetic Kerr parameter $\beta$. They have different
dependence on $r$ as $1/r$ and $1/ r^3$, respectively. It is meant
that near the surface of the star the gravitomagnetic term is in
concurent with  the gravitoelectric one. But in the distance from
the surface of the star which is comparable with its radius $R$
the gravitomagnetic term is ignorable small.

 We plot $F_{1}(\bar r)$ and $F_{2}(\bar r)$ for $\beta=0.1$. The dependence of
these functions on $\bar r$ is shown in Fig.[1] and Fig[2],
respectively. The Goldreich-Julian charge density under general
relativity is shown in Fig.[3].

By least-squares fitting of the curve at $\theta=0$, we find that
Goldreich-Julian charge density decays with the distance away rom
the star as follows:
\begin{equation}
\rho_{GJ}=\frac{10.5053}{r^3}-\frac{5.05692}{r^2}+\frac{1.06093}{r}-
0.084179\ . \nonumber
\end{equation}

The charge density is maximum at the polar cap region and it
remains almost same in certatin extented region in the pole. It is
to be noted that the expression for $\rho_{GJ}$ obtained by
Muslimov and Harding (1997) shows the similar results.

 \begin{center}
 \epsfysize=65mm
 \epsfxsize=95mm
 \epsfbox{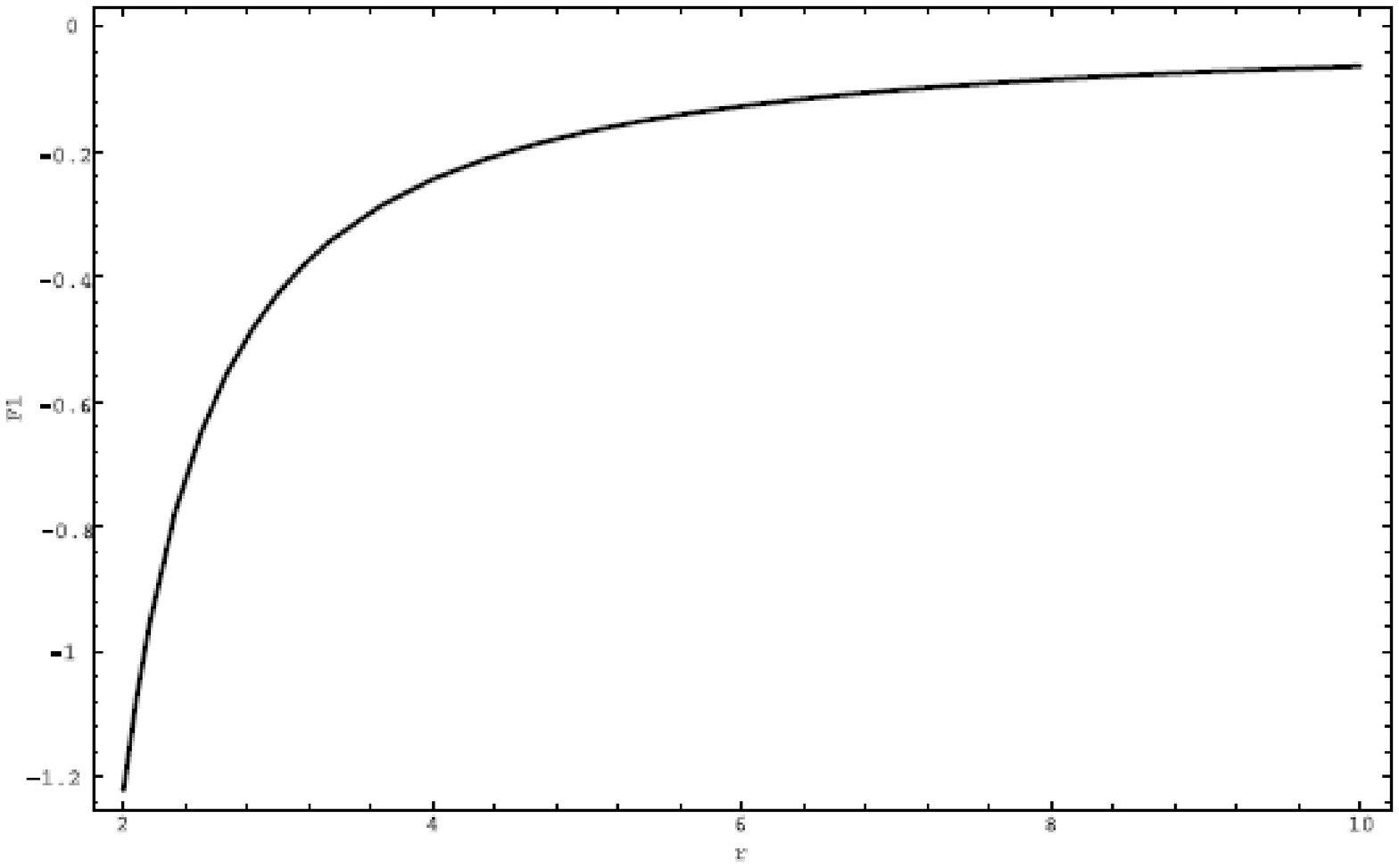}
 \end{center}

Fig.[1]. Goldreich-Julian charge density in general relativity; $
F_{1}$ as function of $\bar r$.

\begin{center}
\epsfysize=65mm
 \epsfxsize=95mm
\epsfbox{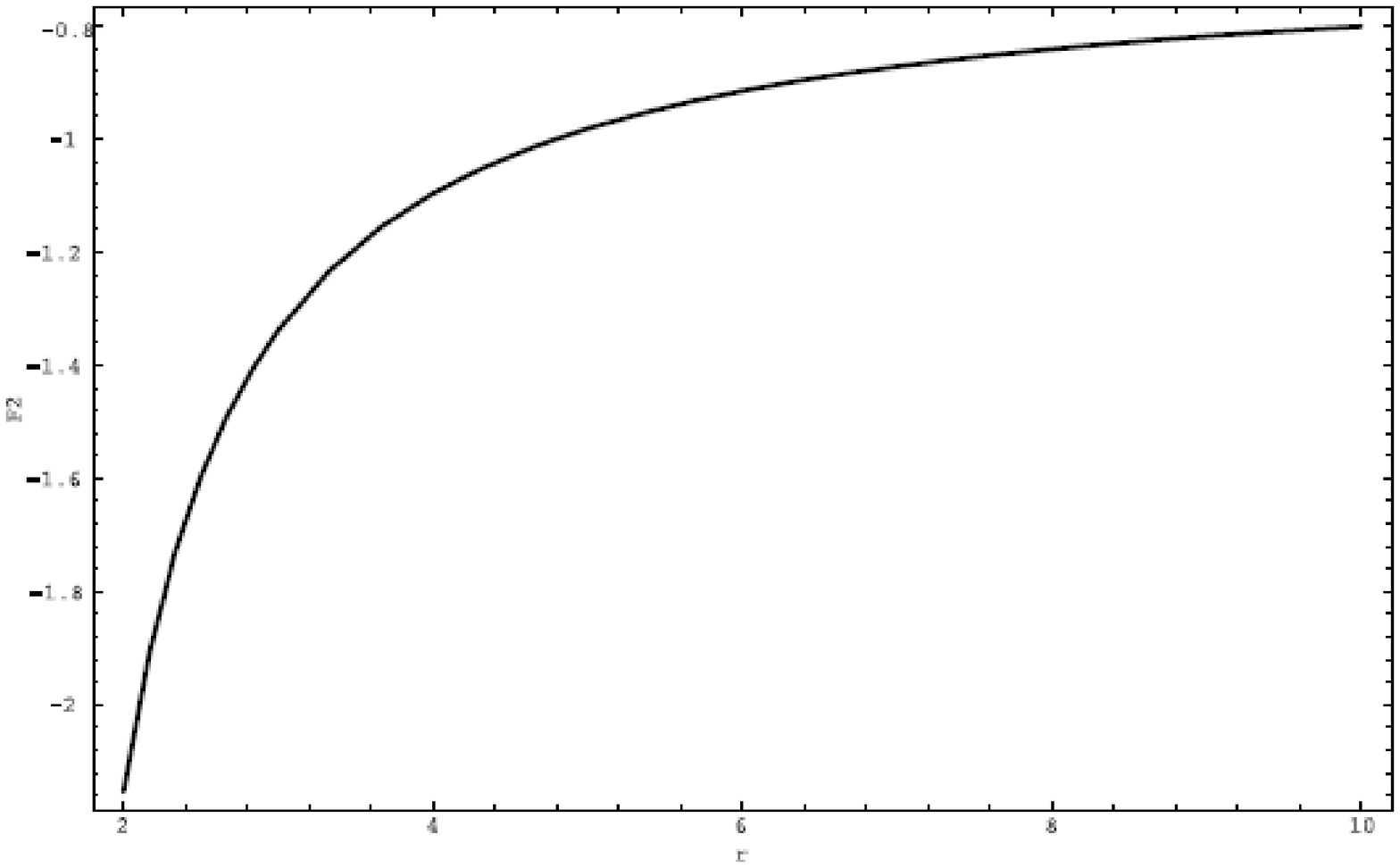}
\end{center}

Fig.[2]. Goldreich-Julian charge density in general relativity;
$F_{2}$ as a function of $\bar r$ .

\begin{center}
\epsfysize=65mm
 \epsfxsize=95mm
\epsfbox{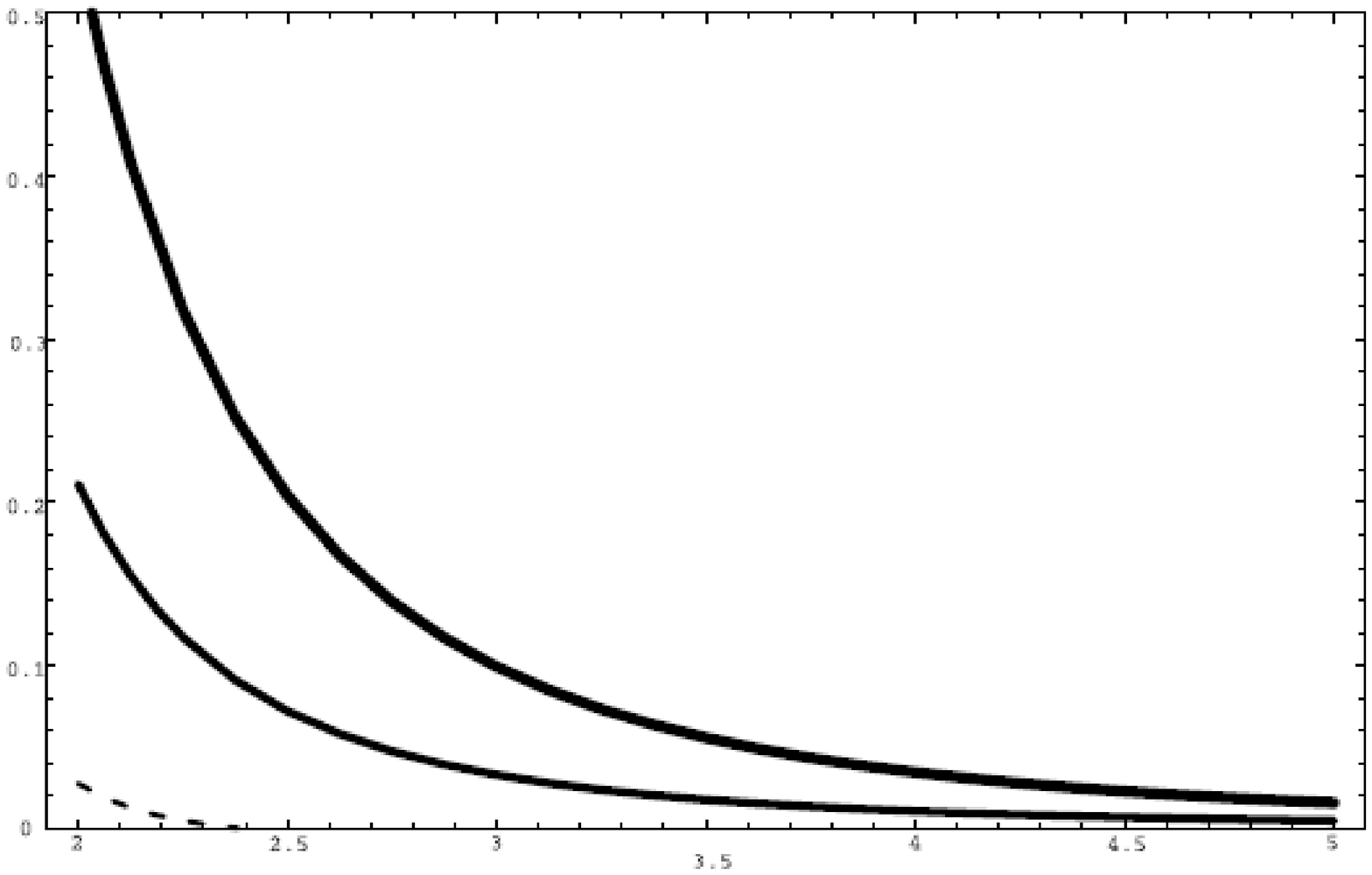}
\end{center}

Fig.[3]. Goldrech-Julian charge density in general relativity
$\rho_{GJ}$ as function of $\bar r$. Thick line corresponds to
$\theta=0$, thin line to $\theta=\pi/4$, and broken line to
$\theta=\pi/3$.

\section{LINEAR PLASMA MODES ALONG THE OPEN FIELD LINES}

 The theory of cascade generation of electron-positron plasma at the polar
cap region of a rotating plasma is developed by Ruderman  \&
Sutherland (1975). According to the theoretical model, due to
escape of charge particles along the open field lines , a polar
potential gap is produced which continuously breaks down by
forming electron positron pair on a time scale of a few
microseconds. A photon of energy greater than $2mc^2$ produces an
electron-positron pair. The electric field of the gap accelerates
the positron out of the gap and accelerates the electron towards
the stellar surface. The electron moves along a curved magnetic
field line and radiate an energetic photon which goes on to
produce a pair as it has a sufficient component of momentum
perpendicular to the magnetic field. Recently, Zhy \& Ruderman
(1997) explained the $e-e^{+}$ pair production from a Crab-like
pulsar. Electrons and positrons are accelerated in opposite
directions to extremely high energies. The Lorentz factor $\gamma$
of the primary electron and positron is given by
$$
e{\mathbf{E}}\cdot{\mathbf{B}}c\approx
\frac{e^{2}}{c^{3}}\gamma^4\left( \frac{c^{2}}{r_c}\right)^2\ ,
$$
where
 This cascade of pair production, acceleration of electrons and positrons
along curved field lines, curvature radiation-pair production
results
 in a "spark" break down of the gap.

 Assuming a steady state thermodynamically equilibrium plasma state in the
polar cap region, we study the linear plasma modes along the field lines.
From the system of equations (10), (6) and (12), we derive the following
linearized equations:

 $${\mathbf{\nabla}}\cdot (\frac{1}{\sqrt{1-\frac{r_{g}}{r}}}{\mathbf{\nabla}} \Phi)=
 -4\pi(\sum_{s}q_{s}\delta n_{s} -\rho_{GJ}) \eqno(20)$$

 $$\frac{\partial}{\partial t'} (\frac{\delta n_{s}}{n_{0}}) + {\mathbf{\nabla}}\cdot (
 \sqrt{1-{{r_{g}}\over r}}{\bf v}_{s}) =0, \eqno(21)$$

 $$\frac{\partial {\bf v_{s}}}{\partial t'} =- \frac{q_{s}}{m} {\mathbf{\nabla}}\Phi,
\eqno(22)$$
 where ${\partial}/{\partial t'} ={\partial}/{\partial t} +
 \left(\frac{\kappa}{r^3} -1\right)\Omega_0{\mathbf{m}}\cdot {\mathbf{\nabla}} $
is the global time derivative along ZAMO trajectories, $ s (=e,
e^{+})$ is the plasma species, $\delta n_{s}$ is the density
fluctuation of the plasma species and $n_{0}$ is the equilibrium
plasma density and $\rho_{GJ}$ is the Goldreich-Julian charge
density as defined by the equation (17).
 The system of equations (20)-(22) is equivalent to the following
 equation:
 $$\frac{\partial}{\partial t^{'2}}\left({\mathbf{\nabla}}\cdot(\frac{1}
{\sqrt{1-{{r_{g}}\over{r}}}}{\mathbf{\nabla}}\Phi)-4\pi\rho_{GJ}\right)+
{\mathbf{\nabla}}\cdot(\omega_{p_{0}}\sqrt{1-\frac{r_{g}}
 {r}}{\mathbf{\nabla}}\Phi)=0, \eqno(23)$$
 where $\omega_{p_{0}}^2 ={8\pi n_{0}e^2}/{m}$.

 Now, by defining the electric field arising from charge separation and the
 corotational electric field, which is the source of $\rho_{GJ}$
  $$ {\bf E} \equiv -\frac{1}{\sqrt{1-\frac{r_{g}}{r}}}{\mathbf{\nabla}}\Phi,\quad
\rho_{GJ}=
 \frac{1}{4\pi}{\mathbf{\nabla}}\cdot{\bf E}_{c}, \eqno(24) $$
 from equation (23), we find

 $$\frac{\partial}{\partial t^{'2}}\large \left[ {\mathbf{\nabla}}\cdot ({\bf E +
E}_{c})\large\right] +
 {\mathbf{\nabla}}\cdot \left[\omega_{po}^2 (1-\frac{r_{g}}{r}) {\bf E} \right]=0, \eqno(25)$$
 which gives
 $$ \frac{\partial}{\partial t^{'2}}{\bf E} + \omega_{po}^2 \left(1-\frac{r_{g}}{r}\right)
 {\bf E}  = - \frac{\partial {\bf E}_{c}}{\partial t^{'2}}. \eqno(26) $$

 For $\frac{\partial {\bf E_{c}}}{\partial t^{'2}} =0$ we may write the
solution of equation (26) as
$$ E = E_{0} \exp\left\{-i\omega_{po}\sqrt{1-\frac{r_{g}}{r}} t'\right\}. \eqno(26)$$

 From the above solution, we find that the plasma frequency in general
relativity  now is defined as
  $$ \omega_{p}^2=\omega_{p_{0}}^2 \left(1-\frac{r_{g}}{r}\right) \eqno(28)$$
  which is equivalent to the gravitational redshift of the oscillation.
  Fig.[4 ] shows the dependence of plasma oscillation on the distance away
from the gravitational radius of the star.

\begin{center}
\epsfysize=65mm
 \epsfxsize=95mm
 \epsfbox{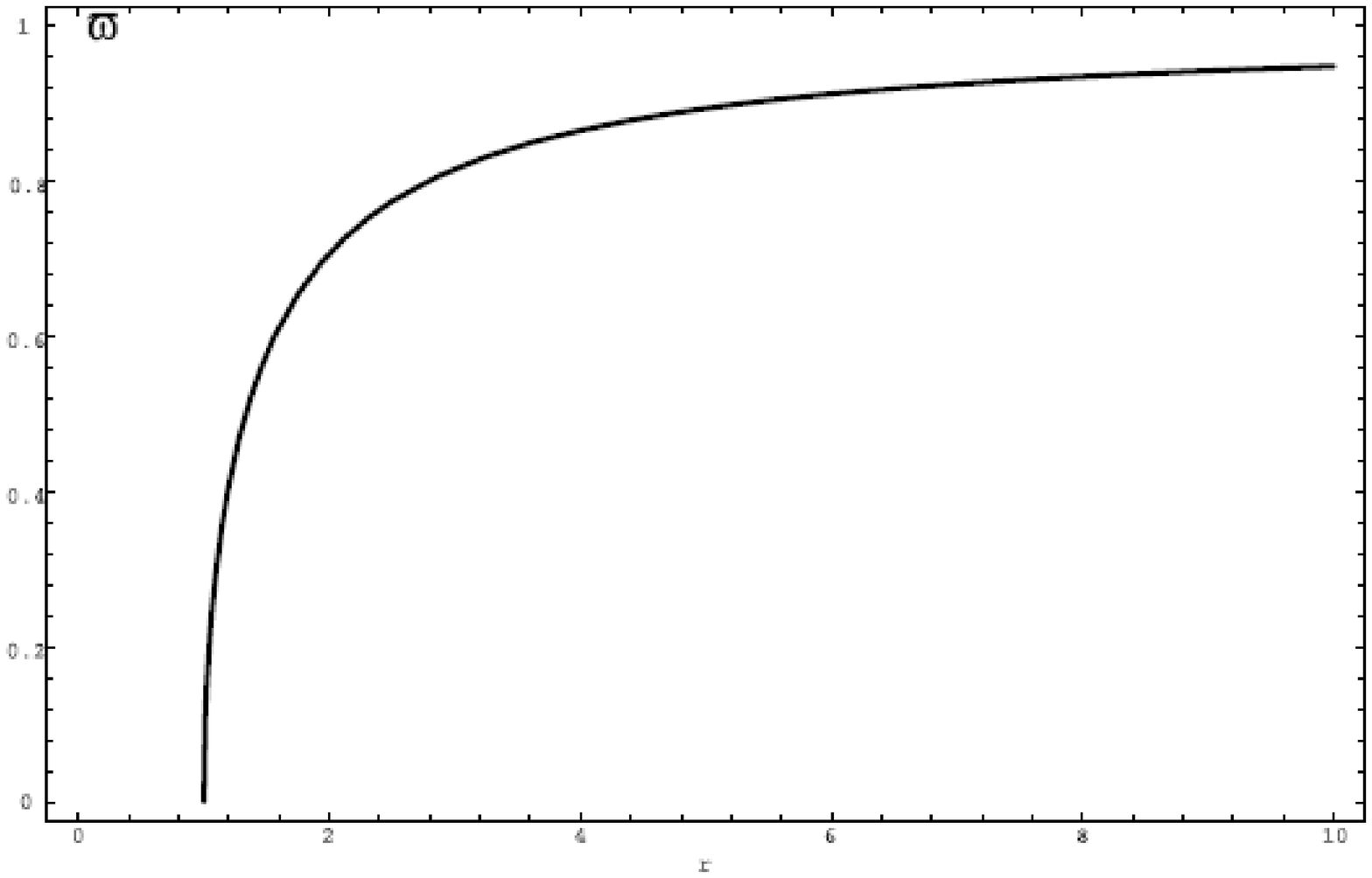}
\end{center}

 Fig.[4]. Plasma frequency $\omega_{p} (r)$ as a function of
 ${\bar{r}}$; gravitational redshift near the Schwarzschild
 radius.

The global time derivative along ZAMO trajectories is defined as
$$\frac{\partial}{\partial t'} = \frac{\partial}{\partial t} +
\left(\frac{\kappa}{r^3}-1\right)\Omega_{0}\frac{\partial}{\partial\phi}.
\eqno(29)$$

Thus, we may define
$$ t' = t +\frac{\phi}{(\frac{\kappa}{r^3}-1)\Omega_{0}}, \eqno(30)$$
 and hence the solution of linear plasma mode is
 $$ E(t,r,\phi) = E_{0} \exp\left\{-i\omega_{p_{0}}{\sqrt{1-\frac{r_{g}}{r}}}( t+
 \frac{\phi}{(\frac{\kappa}{r^3}-1)\Omega_{0}})\right\} \eqno(31)$$

Introducing the dimensionless quantities
 $ {\epsilon} =\frac{ E}{ E_{0}},  \tau= \omega_{po}t, x= \frac{r}{r_{g}},
 \chi = \frac{\omega_{p_{0}}}{\Omega_{0}},\delta = \frac{\kappa}{{r_{g}}^3}$
 from the equation (31), we find
 $$ {\large\epsilon}(\tau,x,\phi)= \sin\left[\sqrt{1-{1\over x}} (\tau +
\frac{\chi}
 {\frac{\delta}{x^3} -1} \phi)\right] \eqno(32). $$

\begin{center}
\epsfysize=65mm
 \epsfxsize=95mm
 \epsfbox{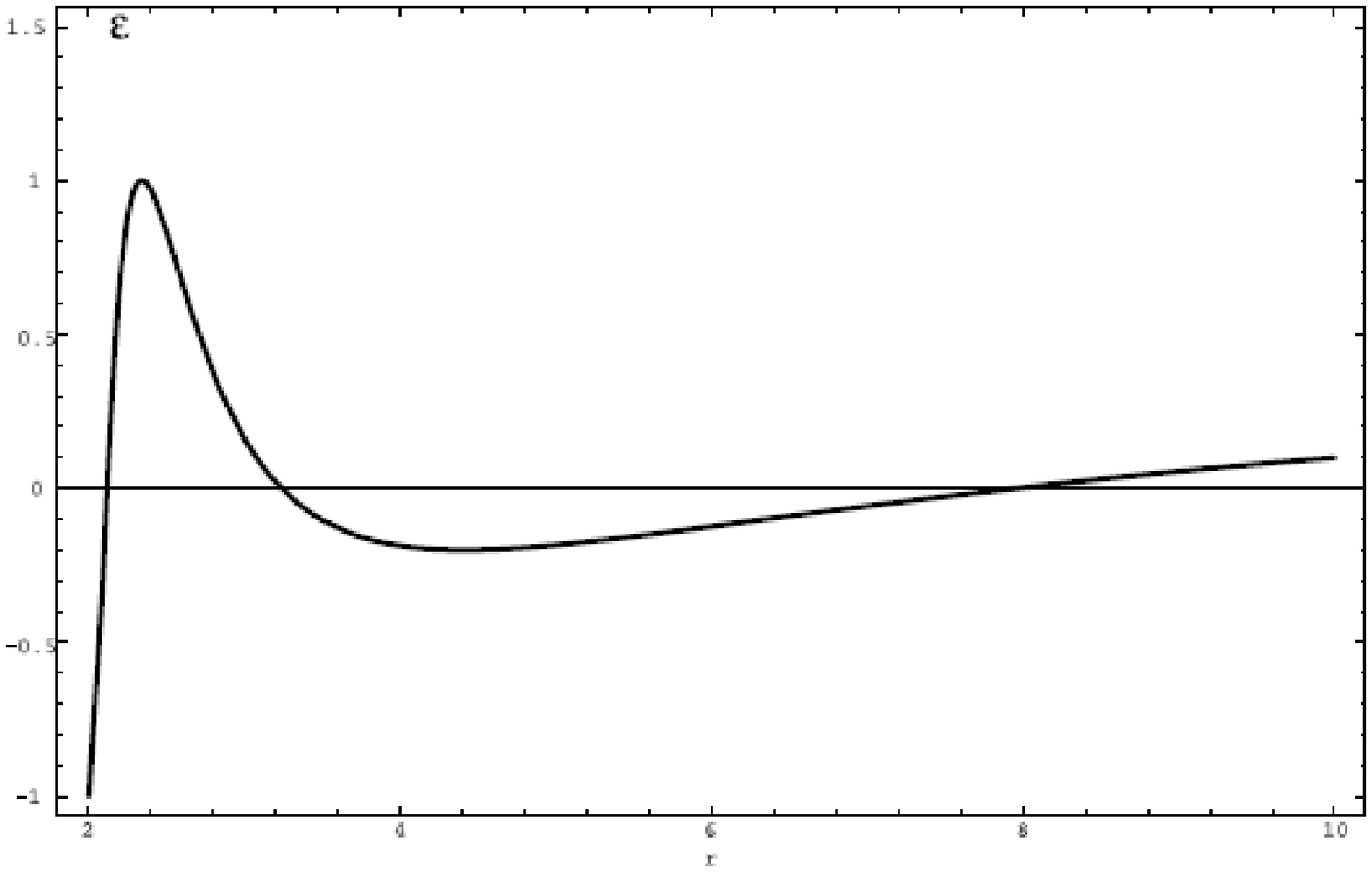}
\end{center}

 Fig.[5]. Linear plasma mode in general relativity; electrostatic
field $\epsilon (\tau , \phi)$ as a function of $\bar{r}$.

 We do some analysis of the linear electrostatic modes around a rotating
neutron star.  First, we consider the Schwarzschild radius equal
to half the radius of the neutron star, i.e.,  $r_{g}=R/2$. Then
we find $\delta =8/5$. Considering relatively dense plasma, we put
$\chi=10^8$. We consider a fixed azimuthal angle $\phi=\pi/4$. For
$\tau=0$ we plot the field $\epsilon(\bar{r})$, which is shown in
Figure 5. We find that the electrostatic field generated by
Goldreich-Julian charge density is maximum near the star surface
and falls quickly from the star.

\section{NONLINEAR PLASMA MODES ALONG THE FIELD LINES}

 Now, we consider nonlinear plasma modes along the open field lines around a
rotating neutron star. The system of equations governing the nonlinear
modes can be written as
 $$\frac{\partial}{\partial t'} (n_{s}) +{\mathbf{\nabla}}\cdot({\sqrt{1-\frac{r_{g}}{r}}}
 n_{s}{\bf v}_{s})=0, \eqno(33)$$

$$\left(\frac{\partial}{\partial t'} + {\sqrt{1-\frac{r_{g}}{r}}}
 \bf v_{s}\cdot{\mathbf{\nabla}}\right)
 {\bf v_{s}}=-\frac{q_{s}}{m} {\mathbf{\nabla}} \Phi, \eqno(34)$$

 $${\mathbf{\nabla}}\cdot(\frac{1}{\sqrt{1-\frac{r_{g}}{r}}}{\mathbf{\nabla}}\Phi)=-4\pi
(\sum_{s}n_{s} q_{s}-\rho_{GJ})\ .\eqno(35)$$

 For the simplicity, we consider ${\bf v_{s}}.{\mathbf{\nabla}} \approx
v_{sr}\cdot{\mathbf{\nabla}}{\partial}/{\partial r}$ (i.e. one
dimensional wave propagation along r) and introduce a moving frame
$\eta= r - Vt',$ where V is a constant. In the considered moving
frame, from equations
 (33) and (34), we get

 $$ n_{s}=\frac{n_{0}V}{V-{\sqrt{1-\frac{r_{g}}{\eta}}}v_{s\eta}}, \eqno(36)$$

$$v_{s\eta}=\frac{1}{mV}(q_{s}\Phi+{\sqrt{1-\frac{r_{g}}{\eta}}}\frac{e^2}{
2mV^2}\Phi^2). \eqno(37)$$

Using equations (36) and (37), in equation (35), we derive the nonlinear
equation for the plasma mode along the field line of the rotating neutron
star:

$$\frac{d^2 \Phi}{d\eta^2} -
\frac{r_{g}}{2\eta^2(1-\frac{r_{g}}{\eta})}\frac{d\Phi} {d\eta}
+\frac{\omega_{po}^2 (1-\frac{r_g}{r})}{V^2} \frac{\Phi}{1-2(1-
\frac{r_{g}}{\eta})(\frac{e\Phi}{mV^2})^2 +{1\over
4}(1-\frac{r_{g}}{\eta})^2 (\frac{e\Phi}{mV^2})^4}$$
 $$=4\pi{\sqrt{1-\frac{r_{g}}{\eta}}}\rho_{GJ}\ . \eqno(38)$$

Now introducing dimensionless quantities

$$\Phi= \frac{e\Phi}{mv^2},\qquad \bar\eta= \frac{\eta}{r_{g}},\qquad
\qquad \bar\omega_{p_{0}}= \frac{\omega_{p_{0}}r_{g}}{V}\eqno(39)$$
we write the equation (38) in dimensionless form:
$$\frac{d^2\Phi}{d\bar\eta^2} - \frac{1}{2\eta(\bar\eta-1)}\frac{d\Phi}
{d\bar\eta} +\bar\omega_{po}^2(1-{1\over\bar\eta})\frac{\Phi}
{1-2(1-{1\over\bar\eta})\Phi^2 +
{1\over 4}(1-{1\over\bar\eta})^2\Phi^4} = F_{c}. \eqno(40)$$
where
$$ F_{c}=-\frac{3\Omega_0\Omega_{c}r_g^2}{2V^2}\left(\frac{R}{\eta}\right)^3
\sqrt{1-\frac{1}{\bar\eta}} \left[ F_{1}(\bar\eta)  \sin^{2}\theta
-F_{2}(\bar\eta)(\sin^2\theta-2\cos^2\theta)\right]. \eqno(41)$$
Here $ \Omega_{c}={eB_{0}}/{mc}$;  $F_{1}(\bar\eta)$ and $
F_{2}(\bar\eta)$ are determined by the equations (18) and (19),
respectively.

\begin{center}
\epsfysize=65mm
 \epsfxsize=95mm
\epsfbox{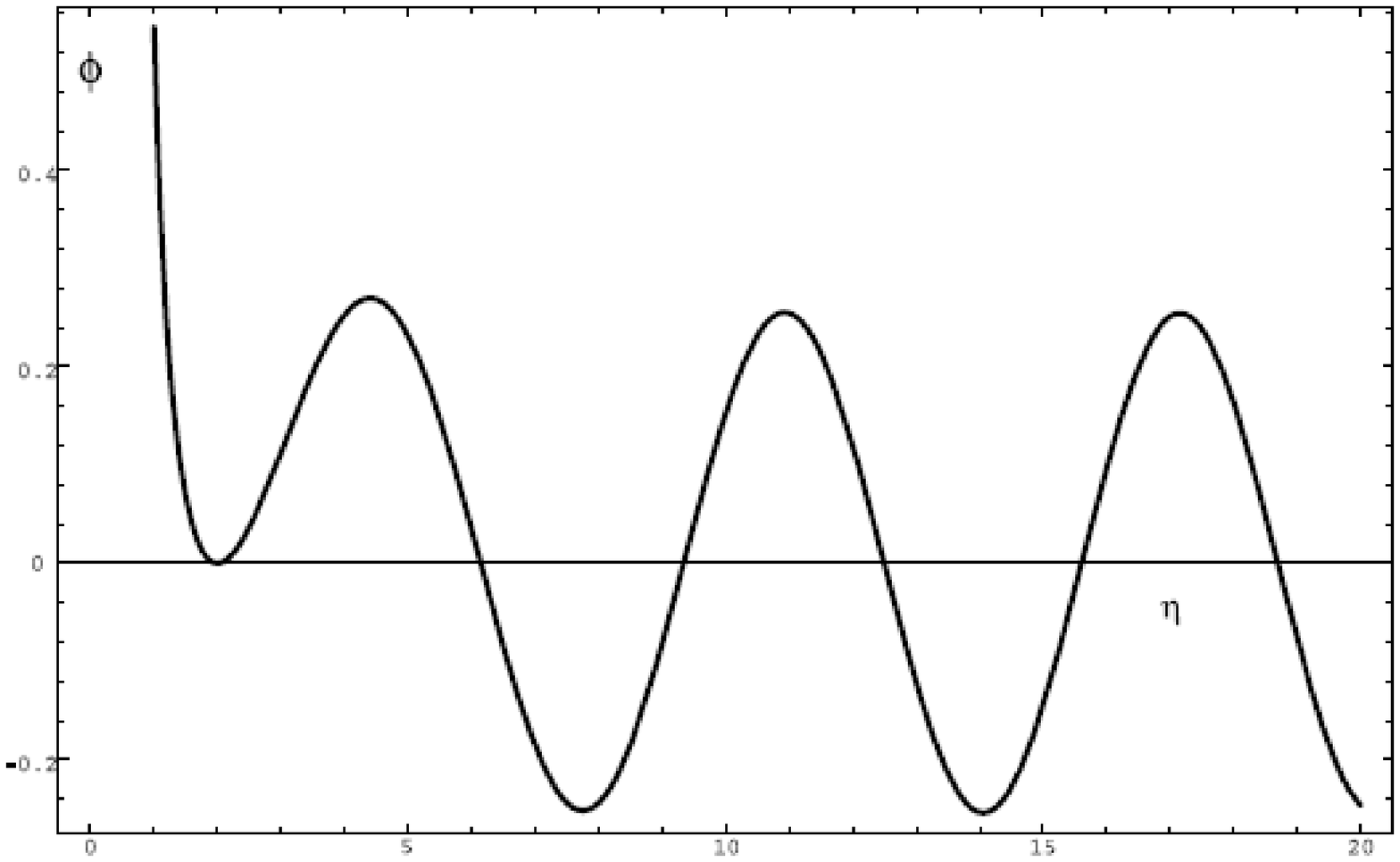}
\end{center}

Fig.[6]. Nonlinear plasma mode in general relativity; propagation
of plasma oscillation $\Phi (\bar\eta)$ near the surface of a
neutron star; numerical solution of the equation (40) with
boundary condition $\Phi (2)=0, $ $\Phi'(2)=0$.

We numerically solve the equation (40) in the polar cap region
$(\theta\approx 0)$ of a neutron star subject to the appropriate
boundary conditions. Following Goldreich $\&$ Julian (1969) and
Muslimov \& Harding (1997), we assume that the surface of a polar
cap and that formed by the last open field lines can be treated as
electric equipotentials. We therefore adopt the condition
$\Phi(r=R)=0$. Second, we require that the steady state component
of electric field parallel to magnetic field vanishes at the polar
cap surface, i.e. $ {d\Phi (r=R)}/{ d\eta}=0$ . By considering
$r_{g}= {R/ 2}$ for a neutron star, we write the boundary
conditions as: $\Phi(2)=0$ and $\Phi'(2)=0 $. The solution of the
equation (40) with the mentioned boundary condition is shown
graphically in Fig.[6]. We find that Goldrech-Julian charge
density creates the initial potential on the surface. Near the
radius the potential is enhanced and it propagates
 almost without damping  along the field lines.

\section{DISCUSSION AND CONCLUSION}

We study the electrostatic plasma modes along the open field lines
of a rotating neutron star.  The dragging of inertial frame and
the effect of general relativity is fully considered in this
study. We perform a detailed analysis of Goldreich-Julian charge
density in general relativity. Since pulsars having smaller
obliquity have larger accelerating drops and this favored for
$\gamma$ -ray pulsar emissions (Muslimov 1995) and it supports the
single pole $\gamma$ - ray pulsar models (Daugherty $\&$ Harding
1994, 1996; Dermer $\&$ Sterner 1994), we confine our analysis in
the zero inclination of the rotating neutron star. As pulsar
radiation takes place in the plasma environment or the radiation
passes through a plasma media , we consider the electrostatic
plasma modes along the open field lines. We study both the linear
and nonlinear modes in the neutron star or black hole plasma
environment. Our general conclusion from the above analysis may be
summarized as follows:

1. Goldreich-Julian charge density is maximum in the polar cap
region and remains almost same in a certain extended  region of
the pole. The charge density exponentially decays with the
distance away from the surface of the star.

2. Plasma oscillation along the field lines resembles to the
gravitational redshift near Schwarzschild radius.

3. Plasma modes grows near the gravitational radius in the black
hole environment.

4. For initially zero field on the surface of a rotating neutron
star, Goldreich-Julian charge density, which is enhanced near the
surface and propagates almost without damping along the open field
lines,  creates the plasma modes.

For further study of plasma dynamics in the neutron star or black
environment we plan to extend our earlier investigations on
solitons (see, Mofiz 1989, 1990, 1993 $\&$ Mofiz et al. 1985,
1995)  propagation along the open field lines of strongly
gravitating objects.

The authors acknowledge the financial support and hospitality at
the Abdus Salam International Centre for Theoretical Physics where
the work was done. Research of BJA is supported in part by the
UzFFR (project 01-06), projects F2.1.09, F2.2.06 and A13-226 of
the UzCST, by the ICTP through the OEA-PRJ-29 and the Regular
Associateship grants and by NATO through the reintegration grant
EAP.RIG.981259.

\section*{REFERENCES}

Arons, J. 1991, in IAU Colloq. 128, The magnetospheric Structure and
Emission
Mechanisms of Radio Pulsars, ed. T. H. Hankins, J. M. Rankin, $\&$ J. A.
 Gil ( Zielona Gora: Pedagogical Univ. Press), 59 \\
Arons, J., $\&$ Scharlemann, E. T. 1979, ApJ,${\bf231}$,854\\
Beskin, V. S. 1990, Soviet Astron. Lett., 16, 286\\
Buzzi, V., $\&$ Hines, K.C. 1995 Phys.Rev.D, $\bf 51$, 6692\\
Daugherty, J. K., $\&$ Harding, A. K. 1994, ApJ, $\bf 429$, 325\\
------------------------------------. 1996, ApJ, $\bf 458$ 278\\
Dermer, C. D., $\&$ Sterner, S. J. 1994, ApJ, $\bf 420$, L75\\
Ginzburg, V. L., $\&$ Ozernoy, L. M. 1964, Zh. Eksp. Teor. Fiz., $\bf 47$,
1030\\
Goldreich, P., $\&$ Julian, W. H. 1969, ApJ, $\bf 157$, 869\\
Landau, L. D.,$\&$ Lifshitz, E. M. 1975, The Classical Theory of Fields,
(Oxford: Pergamon)\\
Mestel, L. 1971, Nature, $\bf 233$, 149\\
---------. 1992, Philos. Trans. R. Soc. London, A, 341, 93\\
Michel, F. C. 1991, Theory of Neutron Star Magnetosphere, ( Chicago:
Univ. Chicago Press)\\
Misner, C. W., Thorne, K. S.,$\&$ Wheeler, J. A. 1973, Gravitation
(San Francisco: Freeman)\\
Mofiz, U. A. 1989, Phys. Rev. A, $\bf 40$, 2203\\
-----------. 1989, Phys. Rev. A, $\bf 40$, 6752\\
-----------. 1990, Phys. Rev. A, $\bf 42$, 960\\
-----------. 1997, Phys. Rev. E, $\bf 55$, 5894\\
Mofiz, U. A., De Angelis, U. $\&$ Forlani, A. 1985, Phys. Rev. A, $\bf 31$,
951\\
Mofiz, U. A., Tsintsadze, L. N. $\&$ Tsintsadze, N. L. 1995, Physica Scri.
$\bf 51$,390\\
Muslimov, A. 1995, Milisecond Pulsars: A Decade of Surprise, ASP Conf.
 Ser. 72, ed. A. S. Fruchter, M. Tavani, $\&$ D. C. Backer ( San Francisco:
ASP), 334\\
Muslimov, A. $\&$ Harding, A. K. 1997, ApJ, $\bf 485$, 735\\
Muslimov, A.,$\&$ Tsygan, A. I. 1986, AZh, 63, 958\\
Ruderman, M.,$\&$ Sutherland, P. G. 1975, ApJ, $\bf 196$, 51\\
Sturrock, P. A. 1971, ApJ, $\bf 164$, 179\\
Thorne, K. S.,Price, R. H., $\&$ Macdonald D. A. 1986, Black Holes: The
Membrane Paradigm (New Haven: Yale Univ. Press)\\
Wasserman, I. M., $\&$ Shapiro, S. L. 1983, ApJ, $\bf 265$, 1036\\
Zhu, T., \& Ruderman, M. 1997, ApJ, 478, 701\\
\end{document}